\documentclass{aastex701}

\begin{document}

\title{The Spitzer Spectroscopic Data Fusion - Merged Spectroscopic Redshift Catalogs in Spitzer Fields}

\author[orcid=0000-0002-6748-0577,sname='Vaccari']{Mattia Vaccari}
\affiliation{Inter-University Institute for Data Intensive Astronomy (IDIA), Department of Astronomy, University of Cape Town, 7701 Rondebosch, Cape Town, South Africa}
\affiliation{University of the Western Cape, 7535 Bellville, Cape Town, South Africa}
\affiliation{INAF - Istituto di Radioastronomia, via Gobetti 101, 40129 Bologna, Italy}
\email[show]{mattia@mattiavaccari.net}  


\begin{abstract}
I present the Spitzer Spectroscopic Data Fusion, a collection of merged spectroscopic redshift catalogs covering fourteen of the most widely studied extragalactic survey fields. Building on the Spitzer Data Fusion multi-wavelength photometric database, the collection merges several publicly available spectroscopic redshift catalogs within each field using a 1\,arcsec matching radius, delivers a single "best" redshift per source together with provenance and overlap flags, and is available on Zenodo as \dataset[doi:10.5281/zenodo.6368347]{https://doi.org/10.5281/zenodo.6368347}.
The dataset is regularly updated as new spectroscopic surveys are published.  It is intended as a community calibration resource for photometric redshift training, SED fitting, and multi-wavelength cross-identification studies.
\end{abstract}



\section{Introduction}

Spectroscopic redshifts are the bedrock of extragalactic astronomy: they calibrate photometric redshift codes, anchor SED fitting, define training sets for machine-learning classifiers, and provide kinematic membership for galaxy groups and clusters.  However, the literature contains dozens of independent spectroscopic campaigns per popular field, each with its own format, coordinate system, quality flag convention, and degree of completeness.  Assembling a coherent, up-to-date compilation for any given
field requires substantial bookkeeping that is routinely duplicated across research groups. The Spitzer Spectroscopic Data Fusion (SSDF) addresses this problem by providing pre-merged, field-level spectroscopic redshift catalogs that are versioned, citable, and publicly archived.  It is a ``building block'' of the broader Spitzer Data Fusion project \citep{Vaccari2015}\footnote{\url{https://www.mattiavaccari.net/df}}, which delivers far-ultraviolet-to-far-infrared photometry for 4.4 million IRAC-selected sources over 65\,deg$^2$.  I describe the catalog construction strategy, the fields covered, the data model, and the intended science applications, and I provide the Zenodo DOI for direct data access.

\section{Catalog Construction}

\subsection{Fields Covered}

The Spitzer Spectroscopic Data Fusion currently covers fourteen extragalactic fields selected on the basis of their rich multi-wavelength legacy and their prominence in Herschel (HerMES, H-ATLAS), radio (MIGHTEE, EMU), and forthcoming Euclid and Rubin surveys. These include the Cosmic Evolution Survey (COSMOS), XMM Large-Scale Structure (XMM-LSS), Chandra Deep Field South (CDFS), Extended Groth Strip (EGS), ELAIS-N1, ELAIS-N2, ELAIS-S1, Lockman Hole (LH), Boötes, Spitzer Extragalactic First Look Survey (FLS), Hubble Deep Field North (HDFN), Akari Deep Field North (ADFN), Akari Deep Field North (ADFS), Euclid Deep Field South (EDFS) fields\footnote{The ADFN field is also known as Euclid Deep Field North (EDFN). The (extended) CDFS field is also known as Euclid Deep Field Fornax (EDFF). The ADFS and EDFS fields as defined here partly overlap, but are provided separately since the original Spitzer coverage was carried out separately and at different bands.}. Fields were selected to match the Spitzer coverage of the corresponding field, ensuring seamless cross-referencing with the photometric database.

\subsection{Merging Strategy}

For each field, all publicly available spectroscopic redshift catalogs (from NED, VizieR, survey team releases, and observatory archives) are gathered and positionally merged using a matching radius of 1\,arcsec. Sources are listed in the merged catalog according to a ranked priority order (CAT1, CAT2, \ldots, CAT$N$), which is defined separately for each field while keeping NED (CAT1) as the highest priority by convention. The column \texttt{ZBEST} records the redshift from the highest-ranked catalog that provides a measurement for each source, and \texttt{ZFLAG} identifies which catalog supplied \texttt{ZBEST}. Individual catalog redshifts \texttt{Z\_1}, \ldots, \texttt{Z\_N} are retained alongside \texttt{ZBEST}, enabling users to apply field-specific quality criteria or to propagate the highest-quality measurement for their particular science case rather than relying on the default priority order. The binary flag \texttt{ZWHERE} encodes, as a bitmask, in how many and which contributing catalogs each source received a redshift measurement: \texttt{ZWHERE} $= \sum_{k=0}^{N-1} 2^k$ for a source present in all $N$ catalogs.  This allows efficient selection of sources with redundant measurements, useful for assessing systematic offsets between surveys. The column \texttt{ZCLASS} is reserved for future population of redshift quality classes but currently mirrors \texttt{ZFLAG}. I note that \texttt{ZBEST} does not necessarily represent the most scientifically optimal redshift for every use case: the priority ranking places NED (CAT1) first by convention, but NED aggregates heterogeneous measurements of varying quality.  Users performing precision photometric redshift calibration should select redshifts from specific high-quality surveys using the per-catalog columns \texttt{Z\_1}, \ldots, \texttt{Z\_N} and the \texttt{ZFLAG} and \texttt{ZWHERE} columns to identify sources with redundant, internally consistent measurements.

\subsection{Versioning and Updates}

The dataset is maintained as a living resource on Zenodo under the DOI \dataset[doi:10.5281/zenodo.6368347]{https://doi.org/10.5281/zenodo.6368347}
(concept DOI, resolving to the latest version), with the most recent release (version 20 April 2026) archived at \dataset[doi:10.5281/zenodo.19928169]{https://doi.org/10.5281/zenodo.19928169}.
Each release is accompanied by a README file (\texttt{AAAREADME.SPECZ-MERGED})
describing the contributing catalogs, their priority ranking, and any field-specific caveats.

\section{Data Model and Access}

\begin{figure*}[ht!]
\plotone{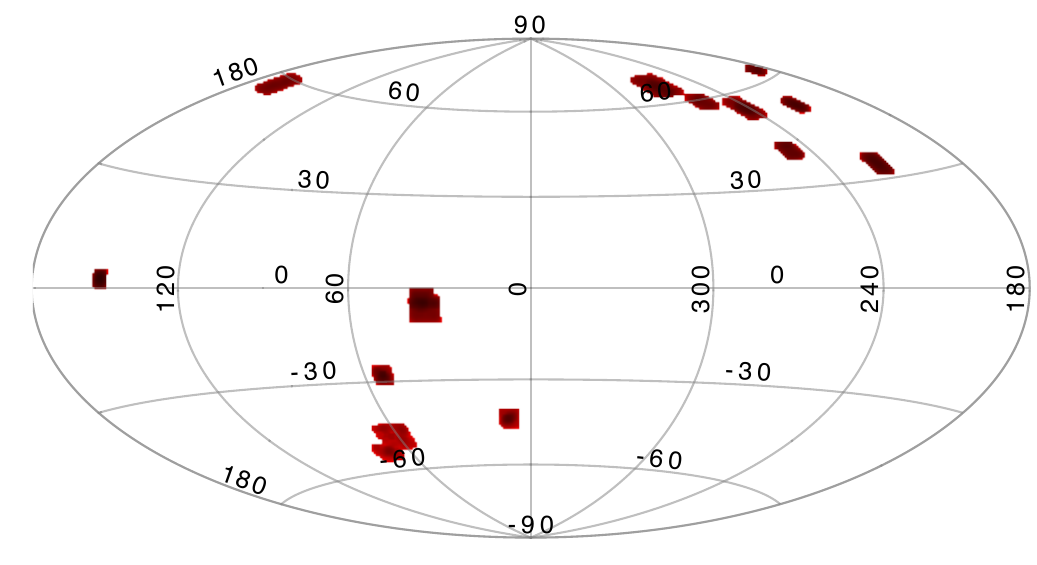}
\caption{Spitzer Spectroscopic Data Fusion Sky Coverage in Equatorial Coordinates.}
\label{fig:fields}
\end{figure*}

All catalogs are distributed as FITS binary tables within a single ZIP archive (310.7\,MB for the 20 April 2026 release).  The archive contains one merged catalog per field, named \texttt{<FIELD>-specz-merged.fits}, together with the README and individual per-survey input catalogs. Figure~\ref{fig:fields} shows the sky coverage of the fourteen fields.

\section{Science Applications}

The Spitzer Spectroscopic Data Fusion has been used in a range of multi-wavelength galaxy evolution studies. Within the Spitzer Data Fusion context, it provided supporting spectroscopic information for luminosity function studies
\citep[e.g.,][]{Vaccari2010,Marchetti2016}.
It also provided training and validation sets for photometric redshift estimation over wide-area infrared-selected samples \citep[e.g.,][]{Pforr2019,RowanRobinson2013}.
In the context of the Herschel Extragalactic Legacy Project
\citep[HELP;][]{Shirley2021}, spectroscopic redshifts compiled within this resource were used to calibrate photometric redshifts for Herschel-selected sources across multiple fields. More recently, the catalogs have served as the reference spectroscopic anchor for multi-wavelength cross-identification in MeerKAT radio continuum surveys, including the MIGHTEE survey \citep{Whittam2024}, where reliable spectroscopic redshifts for radio source counterparts are critical for constructing radio luminosity functions and separating AGN from star-forming galaxy populations. Within the context of ongoing and upcoming Euclid, Rubin and Roman surveys, but also spectroscopic follow-up with e.g. 4MOST, MOONS, PFS and WAVES, the Spitzer Spectroscopic Data Fusion is ideally positioned to provide input spectroscopic catalogs and photometric redshift training and validation sets in the overlapping fields.

\section{Summary}

I present the Spitzer Spectroscopic Data Fusion, a regularly updated,
publicly archived collection of merged spectroscopic redshift catalogs covering fourteen extragalactic survey fields.  The dataset merges heterogeneous spectroscopic campaigns using a uniform 1\,arcsec matching radius and a field-specific priority ranking, and provides per-source provenance and overlap flags to support diverse science use cases.  It is available at \dataset[doi:10.5281/zenodo.6368347]{https://doi.org/10.5281/zenodo.6368347}.
%
\begin{acknowledgments}
I acknowledge support from the European Commission Research Executive Agency (FP7-SPACE-2013-1 GA 607254) and the South African Department of Science and Technology (DST/CON 0134/2014) as part of the Herschel Extragalactic Legacy Project (HELP).  I acknowledge financial support from the Inter-University Institute for Data Intensive Astronomy (IDIA), a partnership of the University of Cape Town, the University of Pretoria, and the University of the Western Cape, and from the South African Department of Science and Innovation's National Research Foundation under the ISARP RADIOMAP Joint Research Scheme (DSI-NRF Grant Number 150551) and the CPRR HIPPO Project (DSI-NRF Grant Number SRUG22031677). This research has made use of the NASA/IPAC Extragalactic Database (NED), which is operated by the Jet Propulsion Laboratory, California Institute of Technology, under contract with the National Aeronautics and Space Administration. I thank the many spectroscopic survey teams whose public data releases make this compilation possible.
\end{acknowledgments}

\facilities{Spitzer(IRAC, MIPS), NED, VizieR, Zenodo}

\software{STILTS/TOPCAT \citep{Taylor2005,Taylor2006}}



\begin{thebibliography}{}

\bibitem[Marchetti et al.(2016)]{Marchetti2016} Marchetti, L., Vaccari, M., Franceschini, A., et al.\ 2016, \mnras, 456, 2, 1999. \doi{10.1093/mnras/stv2717}

\bibitem[Pforr et al.(2019)]{Pforr2019}
Pforr, J., Vaccari, M., Lacy, M., et al.\ 2019,
\mnras, 483, 3168. \doi{10.1093/mnras/sty3075}

\bibitem[Rowan-Robinson et al.(2013)]{RowanRobinson2013} Rowan-Robinson, M., Gonzalez-Solares, E., Vaccari, M., et al.\ 2013, \mnras, 428, 3, 1958. \doi{10.1093/mnras/sts163}

\bibitem[Shirley et al.(2021)]{Shirley2021}
Shirley, R., Duncan, K., Campos Varillas, M. C., et al.\ 2021,
\mnras, 507, 129. \doi{10.1093/mnras/stab1526}

\bibitem[Taylor(2005)]{Taylor2005}
Taylor, M.~B.\ 2005, Astronomical Data Analysis Software and Systems XIV, 347, 29.

\bibitem[Taylor(2006)]{Taylor2006}
Taylor, M.~B.\ 2006, Astronomical Data Analysis Software and Systems XV, 351, 666. 

\bibitem[Vaccari et al.(2010)]{Vaccari2010} Vaccari, M., Marchetti, L., Franceschini, A., et al.\ 2010, \aap, 518, L20. \doi{10.1051/0004-6361/201014694}

\bibitem[Vaccari (2015)]{Vaccari2015}
Vaccari, M.\ 2015, PoS (EXTRA-RADSUR2015), 027.
\doi{10.22323/1.267.0027}

\bibitem[Whittam et al.(2024)]{Whittam2024}
Whittam, I.~H., Prescott, M., Hale, C.~L., et al.\ 2024,
\mnras, 527, 3231. \doi{10.1093/mnras/stad3307}

\end{thebibliography}
\end{document}